\documentclass{article}
\usepackage{amsfonts,amssymb,amsmath}
\usepackage[dvips]{epsfig}
\usepackage[T1]{fontenc}
\usepackage[latin1]{inputenc}
\usepackage{graphicx}
\usepackage[english]{babel}
\usepackage{amsmath}
\usepackage{amssymb}
\usepackage{amsfonts}
\usepackage{color}

\begin{document}


\title{\bf Entropic Considerations on the  Universe and  Universe-Black Hole
Systems}
\author{H. Hadi$^{a}$\thanks{email: hamedhadi1388@gmail.com}, 
F. Darabi$^{a}$\thanks{email: f.darabi@azaruniv.ac.ir}, and Y. Heydarzade$^{b}$\thanks{email: yheydarzade@bilkent.edu.tr}
\\{\small $^{a}$ Department of Physics, Azarbaijan Shahid Madani University,
53714-161 Tabriz, Iran} 
\\{\small $^{b}$Department of Mathematics, Faculty of Sciences, Bilkent University, 06800 Ankara, Turkey}}

\date{\today} \maketitle
\begin{abstract}
We study the entropic considerations on   the Universe system and   the
Universe-Black hole system, filled  by cosmological
constant or exotic quintessence-like  and phantom-like fields having negative pressure,
using their relevant entropic bounds.  It turns out that for both
systems  these considerations  single out
the cosmological constant, among the  negative
pressure candidate fields, as the viable cosmological field.  
 \end{abstract}

\section{Introduction}

The cosmological constant, as a constant energy density in space and time, is the simplest candidate for  dark energy. Actually, $\Lambda$-Cold Dark Matter $(\Lambda CDM$) model
is consistent with the  current observations, however   this model suffers from
the cosmological constant problem \cite{CCP} and coincidence problem \cite{CP}.
These problems have been tackled by introducing some models of dark energy,  the most relevant of which are Quintessence  and Phantom fields. The quintessence, with the equation of state $-1<\omega_{q}<-\frac{1}{3}$, as a canonical scalar field with a particular potential
can describe the late-time cosmic acceleration \cite{Caldwell}. What is considerable about
the quintessence field is that its equation of state dynamically changes with time.
Quintessence field has a prominent role in cosmological dynamics in the presence
of matter and radiation \cite{fujii,ford,wett,ratra}.  The other candidate
for dark energy is the phantom field, with the equation of state $\omega_{p}<-1$,
which can also describe the current cosmological dynamics
\cite{carroll12,singh12,sami12}.  The equation of state parameter for phantom field, when  approaches a constant value,
results
in a big-rip singularity which itself
is a new problem.   Although all these three models of dark energy are capable of 
predicting an accelerating cosmological dynamics, however, their imperfection 
for a  perfect description of current cosmic dynamics, motivates one to revisit these  dark energy models from a non-cosmic point of view. Here, it is worth to mention that there are also other attempts  suggesting an extension of Einstein's general relativity in order to explain
the current accelerating expansion of the Universe  \cite{corda}. What is known as an extended theory of general relativity
is a semi-classical theory with a modification in the Einstein-Hilbert
action. For instance, this modification is done  by considering  higher order curvature invariant
terms or by adding terms where a scalar field is coupled non minimally to
geometry \cite{capo,faroni}. In this framework, $f(R)$ gravity \cite{odd1} is one of the most popular models addressing both  dark energy and dark matter problems.
In \cite{corda}, the possibility of testing the framework of extended theory
of gravity
using gravitational waves  is anticipated. This may be considered as an important  step forward in the framework of extended theories of gravity because of the recent progresses in the gravitational wave astronomy
through the famous LIGO detections.

Another point of view is the powerful thermodynamical approach and the entropic considerations
for the study of dark energy models. It is well known that in principle the equations of motion can   predict
the future of  time-reversible physical systems, however   in reality,
time-reversibility  is not seen for thermodynamical systems due to the entropic
consideration.  Two relevant examples of such thermodynamical systems in the present study are the Padmanabhan's emergent paradigm and Universe-Black hole system surrounded 
by cosmological fields.  In explicit words, we explore the cosmological fields
by entropic considerations and investigate which cosmological field as candidate
for   dark energy model can be
preferred by such entropic considerations in the study of Universe system
and  Universe-Black hole
system.

The idea
that gravity behaves as an emergent phenomenon is referred to the proposal made by
Sakharov in 1967 \cite{sakharov}. In this proposal,  named as the induced gravity, the spacetime background emerges as
a mean field approximation of some underlying microscopic degrees of freedom,
very similar
to hydrodynamics or continuum elasticity theory from molecular physics \cite{visser}. Current research works on the relation between gravity and thermodynamics  support this point of view \cite{pad},
where the major attention is focused on how the gravitational field equations can be
derived from the
thermodynamical point of view. 

In a pioneering work of Jacobson,   the Einstein field equations were obtained,
using the equivalence
principle and Clausius relation $dQ = TdS$ where $Q$, $T$ and $S$ are the heat, temperature and entropy, respectively
   \cite{jacob}. The key point in this work was to demand that the Clausius relation should hold for all local Rindler
causal horizons with $Q$ and $T$ interpreted as the energy flux
and Unruh temperature, as seen by an accelerated observer located inside the horizon. In this
thermodynamic  approach, the Einstein filed equations appear as the equations of state of spacetime.  The Clausius
relation also arises when one treats the gravitational field equations as an entropy
balance law across a null surface, i.e $S_m = S_{grav}$ \cite{pad2}.

In another work by Padmanabhan, 
as a new  approach to show  a relation between gravity and thermodynamics,
 the gravity was shown not to be a fundamental interaction \cite{pad3}. In
this approach, the Newton's law of gravitation was derived by combining
the equipartition law of energy for the horizon degrees of freedom with the
thermodynamical relation $ S = \frac{E}{2T}$ where $S$, $T$ and
$E$ are  entropy,  horizon temperature and  active gravitational mass, respectively \cite{pad4}.
It was also  argued that the current accelerated expansion of the universe
may be derived from the discrepancy between the surface  and
bulk degrees of freedom through the relation
$\Delta V/\Delta t = N_{\rm sur}-N_{\rm
bulk}$, where $N_{\rm
bulk}$ and   $N_{\rm sur}$ are   the degrees of freedom related
to matter-energy content inside the bulk and surface area, respectively \cite{Pad1}.
These studies magnify the importance of thermodynamic
approach to gravity as well as the corresponding thermodynamical quantities, even for cosmological systems.
The entropy and its bounds play the key roles in theses kinds of studies. 

The application of Bekenstein's bound to
sufficiently small regions of the
Universe can be found in  \cite{Bekhadi,Fishadi,Eashadi,Venhadi,Bakhadi,Kalhadi,Bruhadi}.
Moreover, Fischler and
Susskind have proposed a bound \cite{Fishadi} which  can formulate the
holographic principle \cite{Thhadi,Suhadi,Corhadi}. 
They have carefully exposed the difficulties that arise when such bounds are
pushed beyond their range of
validity \cite{Fishadi,Eashadi,Kalhadi}. 
On the other hand, Bousso has proposed D-bound for systems with cosmological horizon, like asymptotically non-flat Schwarzschild-de Sitter black hole  \cite{zzz}.
Of course, one can look for D-bound for other solutions which are not asymptotically flat and include  a cosmological apparent horizon.
A general proposal was then suggested by Bousso \cite{bousso2} for entropy bound, so called covariant entropy bound, which has been shown
to comply with the Bekenstein's 
entropy bound   and the Fischler and
Susskind bound.

In this paper, we  will consider the system
of ``Universe''  and   the system
of ``Universe-black hole'' filled by some exotic fields, and investigate those fields that can be preferred by imposing
 the relevant entropy bounds on these systems. First, for the system
of ``universe'',  we will apply the covariant entropy bound on the light-like cosmological horizon   and  the entropy bound arising from the Padmanabhan's Emergent Paradigm. These two entropy bounds should be identified on the light-like cosmological horizon. The identification of D-bound and Bekenstein
bound, in the framework of the surrounded Universe-Black hole system, can be considered as a suitable criterion by which one can single out
the exotic cosmological fields in agreement with this criterion and rule out the
other  cosmological fields   in disagreement with this criterion. In doing so, for the system of ``Universe-Black hole'' we will apply the D-bound and  the Bekenstein entropy bound on the black    hole. These two entropy bounds should also be identified on the black    hole.
From the mentioned identifications, we will obtain the preferred exotic fields as the viable cosmological fields.
In
 section 2, we find an entropy bound which is resulted by means of  Padmanabhan's Emergent Paradigm and we attempt to identify the maximum entropy bound  coming from the covariant entropy
conjecture, in one hand, and the entropy of Padmanabhan's Emergent Paradigm
on the other hand for universe system.
In section 3, D-bound and Bekenstein bound are studied for 
the   exotic quintessence-like and phantom-like fields, respectively, and their identifications are investigated for universe-black hole system. Finally, in   section 4, we give our concluding remarks.

\section{Entropy bounds for the Universe system}
 
 The relevant entropy bounds which are of particular importance in the present
study for the Universe system are ``Entropy bound of Padmanabhan's emergent Universe'' and Bousso's ``Covariant entropy bound'', as is described in the following argument.
From  observational evidences we know that the universe is almost flat, namely
$k=0$, for which
the Hubble horizon in Padmanabhan's paradigm  becomes exactly the same as apparent horizon for a flat universe. The covariant entropy bound   is introduced based on
the ``\textit{light-sheets}''(null surfaces) \cite{bousso2}, and on the other
hand, the Hubble horizon in the Padmanabhan's Emergent Paradigm becomes a \textit{``null apparent horizon}'' specifically for a flat universe. Therefore,
the Hubble horizon in the Padmanabhan's Emergent Paradigm for a flat universe
can be considered
as a system for which one can ascribe the covariant entropy bound. 
In other words, if $k=0$ is provided, then the Hubble horizon in  Padmanabhan's paradigm plays the role of  null surface enclosing the Universe  and the
covariant entropy bound becomes applicable to this system. For details of discussion, refer to cosmological corollary of covariant entropy bound in \cite{bousso2}. 
 
 \subsection{  Entropy bound for Padmanabhan's emergent flat Universe}
In this subsection, we show that how Padmanabhan's emergent paradigm
for an expanding flat universe  can be written as an 
 ``emergent lower entropy bound'' for the Universe system.  
 
According to Padmanabhan's proposal, the difference between the
surface degrees of freedom and the bulk degrees of freedom in a
region of space may result in the accelerated expansion of the Friedmann-Robertson-Walker
(FRW) Universe through the relation $\Delta V/\Delta t = N_{\rm sur}-N_{\rm
bulk}$  where $N_{\rm bulk}$ and  $N_{\rm sur}$ are  referred to the degrees of freedom related
to matter and energy content inside the bulk and surface area, respectively \cite{Pad1}.

 For an expanding Universe, we have the following condition  for the Padmanabhan's formula
\begin{equation}\label{a}
\frac{\Delta V}{\Delta t}\geqslant0,
\end{equation}
which demands
\begin{equation}\label{c}
 N_{\rm sur}-N_{\rm
bulk}\geqslant0.
\end{equation}
 On the other hand,
we know that the relation between surface entropy $S_{sur}$ and surface degrees of 
freedom is as follows
\begin{equation}\label{4s}
4S_{sur}=N_{sur},
\end{equation}
 where the entropy of the surface is $\frac{A}{4}$,  $A$ being the area of the surface enclosed by the Hubble horizon $r_{H}$.
One can also write the bulk degrees of freedom in terms of  its  energy $E$
and temperature $T$ as
\begin{equation}\label{N}
N_{bulk}=\frac{2E}{T},
\end{equation}
where the thermodynamic temperature of our cosmological system is
 ${H}/{2\pi}$. So, one can
rewrite the equations (\ref{c}), (\ref{4s}) and (\ref{N}) as follows
\begin{equation}\label{entropy}
\pi r_{H}E\leqslant S_{sur},
\end{equation}
which can be interpreted  as a definition of  the ``emergent lower entropy bound''.  The reason why we call  (\ref{entropy}) as the emergent lower entropy bound is that it is a trivial rewriting of the Friedmann equation in terms of nonstandard variables $r_{H}, E$ and $S$,
and has no independent content. For example, unlike $S$ in the covariant bound, Padmanabhan's $N_{bulk}$ is not defined as the von Neumann entropy or the thermodynamic entropy of an actual bulk matter system, rather it is just a suggestive name given to a quantity that is directly defined in terms of quantities like $H, \rho$, and $p$ that appear in the Friedmann equation. So, there is no nontrivial content to the statement that the Friedmann equation can be expressed in terms of such quantities. That is why the relation (\ref{entropy})
cannot be considered as a definition of a  ``lower entropy bound''
for the surface entropy $S_{sur}$, so it merely  can be interpreted as a definition
of ``emergent lower entropy bound'' of a cosmological system  in the framework of emergent Universe scenario. Therefore, it is
  meaningless to compare in general the ``covariant upper entropy bound'' with ``$\pi r_{H}E$
  as the emergent lower bound of $S_{sur}$'', unless some specific conditions are provided
in order for this comparison becomes meaningful. In the following section, we discuss that there is one  specific possibility providing
a meaningful comparison which corresponds to a spatially flat universe, i.e $k=0$.

 \subsection{Covariant entropy bound for the flat Universe}

In order to apply the covariant entropy bound on the flat Universe, one should
 study the relevant energy contents of the flat Universe, namely Misner-Sharp energy and Komar energy, and  investigate which of these two  is successful in matching  the covariant entropy bound with the entropy bound of Padmanabhan's emergent paradigm, for the flat Universe.
 \subsubsection{Misner-Sharp energy}

Let us start with the Misner- Sharp energy  \cite{mis}
\begin{equation}
E=\int T_{\mu\nu}u^\mu u^\nu dV,
\end{equation}
where $u^\mu =\delta^\mu_0$, $T_{\mu\nu}$ and $V$ are the energy-momentum
source and the volume of the  bulk space, respectively. Then, the total  Misner-Sharp energy inside the Hubble horizon reads as 
\begin{equation}\label{mass}
M(r_H)=\int_0^{r_{H}}4\pi r^2\rho dr=\frac{4\pi}{3}r_{H}^{3}\rho,
\end{equation}
where $r_H$ is the Hubble horizon radius and $M=E$. Moreover, for the apparent
horizon we have $r=2M(r)$  in which for our cosmological case with a flat
spatial geometry the apparent and Hubble horizons coincides and consequently
this formula takes the form of $r_{H}=2M(r_H)$. Also, using the Friedmann
equations for $k=0$, we have $r_{H}=\sqrt \frac{3}{8\pi \rho}$. Then, using
 (\ref{entropy}) and (\ref{mass}), we obtain
\begin{equation}\label{kkk}
\frac{\pi r_{H}^{2}}{2}\leqslant S_{sur}.
\end{equation}
 On the other hand, the maximum of Bousso's covariant entropy bound for $k=0$ and the null surface
 defined by $r_{H}$ is given by 
 \begin{equation}\label{Bb}
 S=\frac{A}{4}=\pi r_{H}^2.
 \end{equation}
We demand that the inequality (\ref{entropy}) to be saturated for $k=0$ and the null surface defined by $r_{H}$  as
\begin{equation}\label{smm}
S_{sur}=\pi r_{H}E,
 \end{equation}
such that it can be compared
with (\ref{Bb}) on the null surface defined by $r_{H}$ . In doing so, if we put the Misner-Sharp energy $E=\frac{r_{H}}{2}$ in (\ref{smm}),
we arrive at the result that the Misner-Sharp Energy has no capability for having the equal values of entropy bounds (\ref{Bb}) and (\ref{smm}) on {\it
the ``null surface of cosmological apparent horizon''}. Therefore, this energy definition fails in matching two entropy
 bounds.

\subsubsection{Komar energy}
In this subsection, we show that Komar energy is capable for removing the
above mentioned inconsistency for the Misner-Sharp energy. To begin with, we consider the Komar energy  \cite{Kom}
\begin{equation}
E=\int \left(2T_{\mu\nu}-Tg_{\mu\nu}\right)u^\mu u^\nu dV,
\end{equation}
where $T$ is the trace of the energy-momentum source $T_{\mu\nu}$. Then, the total Komar
energy
in the bulk space enclosed by the surface of the Hubble horizon is given by \cite{Pad1}
\begin{equation}\label{komar}
E(r_{H})=\int_0^{r_{H}}4\pi r^2|\rho+3p| dr=\frac{4\pi}{3}r_{H}^{3}\rho
|1+3\omega|,
\end{equation}
where we have considered the barotropic equation of state $p=\omega \rho$\footnote{Regarding
that both the degrees of freedom and entropy are positive definite quantities, the absolute value in the integrand in (\ref{komar}) is considered  for the positivity of the bulk degrees of freedom in Eq.(\ref{N}) as well as for the  entropy in Eq.(\ref{relation}). In fact, one has to consider the absolute value of energy or follows Padmanabhan \cite{Pad1} by introducing the $\epsilon$ parameter to assure the positivity of bulk degrees of freedom for both the positive or negative
values of $\rho+3p$, see the Eq.(8) in \cite{Pad1}.}. Then, from the inequality (\ref{entropy}), we obtain
\begin{equation}\label{pad bound}
4\pi r_{H}(\frac{4\pi}{3}r_{H}^{3}\rho)
|1+3\omega|\leqslant A=4S,
\end{equation}
where the L.H.S  becomes maximum (equality case) at $r_H$ as
\begin{equation}\label{*sur}
A=4S_{sur}={2\pi r_{H}(2M(r_{H}))|1+3\omega|}.
\end{equation}
Then, using (\ref{*sur}), we obtain
\begin{equation}\label{relation}
S_{sur}=\frac{\pi r_{H}^{2}|1+3\omega|}{2}.
\end{equation}
This shows that, unlike the Misner-Sharp energy, the Komar energy has capability for having the equal values of entropy bounds (\ref{Bb}) and (\ref{relation}) on the cosmological apparent horizon null surface for two specific values of $\omega$ as $\omega=1/3$ and $\omega=-1$ corresponding to the radiation and de Sitter universes, respectively.

It is interesting to note that without  comparison between the covariant entropy bound
and  entropy bound which comes from Padmanabhan's  emergent paradigm, one cannot
reach to the correct relation  (\ref{relation}) between the entropy of the
enclosing surface and the physical quantities within the bulk space. This relation  indicates the holographic behaviour of the system  in a beautiful
way and shows that the holographic principle demands the Komar energy as
the correct energy content of the cosmological bulk space.  

In the following section, based on
what we obtained, we discuss that one may obtain the current accelerating
expansion of the universe as a matching condition of two cosmological entropy
bounds resulting from the covariant bound and emergent paradigm.

\section{Entropy bounds for the Universe-Black hole system filled
by cosmological
fields}

The relevant entropy bounds which are of particular importance in the present
study for the Universe-Black hole system filled
by cosmological
fields are
D-bound and Bekenstein bound. This is because both the D-bound and Bekenstein bound are applicable  on this system having both event horizon and cosmological
apparent horizon. This feature lets us to compare D-bound with Bekenstein bound for the Universe-Black hole system filled
by cosmological
fields.

The black hole inside the Universe-Black hole system filled
by cosmological
field is considered as a solution surrounded by the cosmological
field. The metric of this solution is
given by \cite{1}
\begin{equation}\label{metricone}
ds^{2}=-\left(1-\frac{r_{g}}{r}-(\frac{r_{s}}{r})^{{3\omega_{s}+1}}\right)dt^{2}+\frac{dr^{2}}{(1-\frac{r_{g}}{r}-(\frac{r_{s}}{r})^{{3\omega_{s}+1}})}+r^2d\Omega^2.
\end{equation}
Note that the metric (\ref{metricone}) is an exact and a general solution for the Universe-Black hole system in the sense that the equation of
state of the surrounding cosmological  field $p_{s}=\omega_{s}\rho_{s}$ is generic. Going through 
\cite{1} which  specifically studies the  exotic quintessence-like field,  one finds that the derivation of the solution is completely general and not specific
to the quintessence-like field;  it  can reduce to its specific limits, like the black hole in the radiation or de Sitter
background, by the particular choices of the equation of state parameter
$\omega_{s}$. The main origin and motivation for the Kiselev solution \cite{1} is  the current
acceleration of the universe. This implies that such a solutions is 
important from a cosmological point of view, like the Schwarzschild-de Sitter
solution, in the sense that it describes a black hole embedded in a cosmological
field governing the current accelerating expansion phase of the universe. This
fact motivated us to investigate this solution as a model for the universe-black
hole system. In Kiselev solution,
each arbitrary choice of $\omega_{s}$ represents an arbitrary cosmological background field, and so the solution (\ref{metricone}) describes a black hole embedded in that cosmological background
field. For example:
\begin{itemize}
\item 
The cases $\omega_{s}=0$ and $\omega_{s}=1/3$ represent the dust and radiation backgrounds,
see for example \cite{majeed}. These particular limits are also mentioned in \cite{1}  at the beginning of section 3, and by following its derivation
one can recognize the generality of this solution.  
\end{itemize}
\begin{itemize}
\item 
By setting $\omega_{s}=-1$, the general solution in \cite{1} reduces to the Schwarzschild-de
Sitter solution, the entropy D-bound of which    is obtained  in \cite{zzz}.
\end{itemize}
\begin{itemize}
\item 
The case $-1<\omega < -1/3$ represents the background  exotic quintessence-like
field with a negative pressure which is a  candidate for  dark energy as the
source for the current accelerating expansion of the universe \cite{hellerman}.
It is worth mentioning that we name this field as ``quintessence-like''
 since in it has no dynamics, and its energy-momentum source is not a perfect fluid.
\end{itemize}
\begin{itemize}
\item 
The range $\omega < -1$ corresponds to exotic phantom-like fields with super-negative pressures
as other candidates for the dark energy \cite{vikman}. The  final fate of a universe 
dominated by a dynamical phantom field is Big Rip \cite{rr}.
The same argument on the name of  quintessence-like  field applies for this case. \end{itemize}

 \subsection{D-bound and Bekenstein bound for the Universe-black hole system
filled
by exotic  quintessence-like
field}

In the  case of  neutral black hole surrounded by quintessence with $ w_{q} = -2/3$
\cite{1} the metric (\ref{metricone}) becomes
\begin{equation}
ds^{2}=-\left(1-\frac{r_{g}}{r}-(\frac{r_{q}}{r})^{-1}\right)dt^{2}+\frac{dr^{2}}{(1-\frac{r_{g}}{r}-(\frac{r_{q}}{r})^{-1})}+r^2d\Omega^2.
\end{equation}
 The inner and outer horizons are obtained by the equation $g^{rr}=0$ as follows
\begin{equation}\label{rin}
r_{in}=\frac{1}{2}(r_{q}-\sqrt{r_{q}^{2}-4r_{q}r_{g}}),
\end{equation}
\begin{equation}
r_{out}=\frac{1}{2}(r_{q}+\sqrt{r_{q}^{2}-4r_{q}r_{g}}),
\end{equation}
where $r_{q}>4r_{g}$ and $r_{g}<r_{in}<r_{out}<r_{q}$. For constructing the 
D-bound \cite{zzz} we take the following procedure.  D-bound is
derived by assuming a matter system inside the apparent cosmological horizon of an observer inside a Universe, with a future de-Sitter asymptotic, which
is
then converted into an  empty pure de-Sitter space
through a thermodynamical process  by which
the matter system is pushed outward  the cosmological horizon.  The total entropy of  the asymptotic de Sitter space, including the matter system, as the initial thermodynamical
system is given by
\begin{equation}
S=S_{m}+\frac{A_{c}}{4},\label{entropyi}
\end{equation}
where $S_{m}$ is the entropy of the initial matter   inside the cosmological
horizon and $A_{c}/4$ is the Bekenstein-Hawking entropy of the  apparent cosmological horizon, enclosing matter system, with surface area $A_{c}$. The final entropy of the system, after matter evacuation, will be $S_{0}=A_{0}/4$
in which $A_{0}$ is the area of cosmological horizon of the
pure de Sitter space. Considering the generalized second law of thermodynamics $S\leq S_{0}$,
we obtain \cite{zzz}
\begin{equation}\label{dbounddd}
S_{m}\leqslant\frac{A_{0}}{4}-\frac{A_{c}}{4},
\end{equation}
which is the so-called \emph{D-bound} on the matter system. 

In the case of Universe-Black hole system filled
by exotic quintessence-like
field, $A_{0}$ and $A_{c}$
are the area of the horizons enclosing  the quintessence-like field and the
quintessence-like plus matter
fields, respectively.  Using 
$r_{0}=r_{out}\mid_{_{m=0}}=r_{q}$ and
$r_{c}=r_{out}$, the D-bound becomes
\begin{equation}
S_{m}\leqslant \pi\left(\frac{r^{2}_{q}}{2}+r_{q}r_{g}-\frac{r_{q}}{2}\sqrt{r_{q}^{2}-4r_{q}r_{g}}\right).
\end{equation}
For $r_{q}\gg r_{g}$,  the D-bound leads to 
\begin{equation}
S_{m}\leqslant 2\pi r_{q}r_{g}.
\end{equation}
On the other hand, the Bekenstein bound is given by \cite{zzz} 
\begin{equation}\label{bekenflat}
S_{m}\leqslant \pi r_{in}R,
\end {equation}
where $r_{in}$ is the gravitational radius of the system and $R=r_{out}$
is the geometric radius of the system. For $r_{q}\gg r_{g}$ the Bekenstein bound becomes
\begin{equation}\label{bek}
S_{m}\leqslant \pi r_{q}r_{g.}
\end{equation}
Therefore, it turns out that in this case the Bekenstein bound and the D-bound are not the same for
dilute system. Indeed, the Bekenstein bound is tighter than the D-bound.
\subsection{D-bound and Bekenstein bound for the Universe-Black hole system
filled
by exotic phantom-like
field}
In the  case of  neutral black hole surrounded by exotic phantom-like
field with $ w_{p} = -4/3$, the metric ($\ref{metricone}$) reads as 
\begin{equation}
ds^{2}=-(1-\frac{r_{g}}{r}-(\frac{r_{p}}{r})^{-3})dt^{2}+\frac{dr^{2}}{(1-\frac{r_{g}}{r}-(\frac{r_{p}}{r})^{-3})}+r^2d\Omega^2.
\end{equation}
Because of $g^{rr}=0$ the inner and outer horizons for $r_{p}\gg r_{g}$
are as follows
\begin{equation}
r_{in}=r_{g}+r_{g}^{4}r_{p}^{-3}+O(r_{p}^{-6}),
\end{equation}
\begin{equation}
r_{out}=r_{p}-\frac{r_{g}}{3}-\frac{2r_{g}}{9r_{p}}-\frac{20}{81}r_{g}^{3}r_{p}^{-2}-\frac{1}{3}r_{g}^{4}r_{p}^{-3}+O(r_{p}^{-4}).
\end{equation}
To derive the D-bound for this system we use the same equation ($\ref{dbounddd}$). In this case
$A_{0}$ and $A_{c}$
are the area of the horizons enclosing  the exotic phantom-like field and the
phantom-like plus matter
fields, respectively. For $r_{p}\gg r_{g}$, using $r_{0}=r_{out}\mid_{_{m=0}}=r_{p}$ and
$r_{c}=r_{out}$, the D-bound
is obtained as
\begin{equation}
 S_{m}\leqslant  \frac{2}{3}\pi r_{g}r_{p}.
\end{equation}
The Bekenstein bound ($\ref{bekenflat}$) for this case is given by
\begin{equation}
S_{m}\leqslant  \pi r_{g}r_{p}.
\end{equation} 
 For this case, the D-bound bound is tighter than the Bekenstein bound. It turns out that, as for the exotic quintessence-like field, for the exotic Phantom-like field
also the Bekenstein bound and the D-bound are not the same for
dilute systems.  Then, regarding the gravitational nature of both the quintessence-like and phantom-like fields in which both of them are violating
the strong energy condition but they lead to the looser and tighter D-bounds relative to Bekenstein bound, a question raises up here. Is there any particular field violating the strong energy condition but providing the same result for both these entropy bounds? In other words, what is the matching condition
for these two bounds? It is shown by Bousso that for the case of cosmological constant, both these entropy bounds coincide \cite{zzz}. Then, regarding our obtained
results for the exotic quintessence-like and phantom-like fields along with \cite{zzz},
it is seen that the D-bound and Bekenstein bound coincide for the Universe-Black hole system  {\it only for the cosmological
constant}. These findings prove that although all the quintessence-like, cosmological
constant and phantom-like fields are motivated for deriving the current accelerating expansion
of the universe, but the cosmological constant is the only
 viable cosmological field from  entropic point of view.   
\section{Discussion and concluding remarks}

By applying Bousso's covariant entropy conjecture for the cosmological spatial region in one hand, and the entropy bound which comes from the Padmanabhan's Emergent Paradigm, on the other hand, we have shown that these two entropy
bounds are in agreement just for the flat ($k=0$) FRW Universe and are equal
to  the maximal entropy on the ``null surface" defined by Hubble horizon $r_{H}$, provided that: 
\begin{itemize}
\item inside of the apparent horizon be filled by the radiation, namely $\omega=\frac{1}{3}$, \end{itemize} 
or
\begin{itemize}
\item inside of the apparent horizon be pure de Sitter space subject to the cosmological constant, namely  $\omega=-1$. 
\end{itemize}
In other words,  the maximal entropy inside the apparent horizon of the flat
FRW universe occurs when it is filled completely just by the radiation field or cosmological constant. However, considering the fact that $\omega=\frac{1}{3}$
case cannot describe the accelerating behavior of the Universe, in the context
of the Padmanabhan's Emergent Paradigm, we can leave this case and just
keep $\omega=-1$ case. We arrive at the conclusion that the cosmological fields with $\omega \neq-1$,
 such as phantom-like and quintessence-like, are ruled out because of non-compatibility
of the covariant entropy  bound and the entropy bound  coming from the Padmanabhan's Emergent Paradigm\footnote{ The weirdness of the phantom  fluid,
which seems to violate the second law of thermodynamics in many ways has been considered in \cite{phy1,phy2,phy3,phy4,phy5,phy6,phy7} and may turn out to be completely
un-physical.}.

 The same result is obtained for the Universe-Black hole system  filled by the cosmological
fields  as follows. We know that D-bound and Bekenstein
bound are the direct results of GSL. Therefore, we conclude that both of them
must lead to the same entropy bound  when imposing on a certain matter
system. The  cosmological constant
 is known to contribute to the metric as $r^{2}$ term and for
this contribution  the D-bound is identified with the Bekenstein bound
in dilute systems, in complete agreement with the above mentioned conclusion. Any deviation from  $r^{2}$ term corresponding to the  contributions of exotic
quintessence-like and phantom-like fields
 as  $r$ and $r^{3}$ terms
leads to D-bounds
looser and tighter than the Bekenstein bound, respectively. Therefore, the quintessence-like and phantom-like fields are ruled out again because of non-agreement
of the D-bound and the Bekenstein bound.
These features turns out to be a consequence of weakness and strongness of repulsion forces
corresponding to quintessence-like and phantomlike fields in comparison to the repulsion
force corresponding to the cosmological constant.   
 


\end{document}